\documentclass[10pt,conference]{IEEEtran}
\IEEEoverridecommandlockouts
\usepackage{booktabs}
\usepackage{multirow}
\usepackage{stfloats}
\usepackage{cite}
\usepackage{amsmath,amssymb,amsfonts}
\usepackage{algorithmic}
\usepackage{graphicx}
\usepackage{textcomp}
\usepackage{xcolor}

\def\BibTeX{{\rm B\kern-.05em{\sc i\kern-.025em b}\kern-.08em
    T\kern-.1667em\lower.7ex\hbox{E}\kern-.125emX}}

\usepackage{marvosym}  
    
\begin{document}
\title{Generalizable Audio-Visual Navigation via Binaural Difference Attention and Action Transition Prediction}
\author{
Jia Li$^{1,2,3}$, and Yinfeng Yu$^{1,2,3}$$^{,\mbox{\Letter}}$%
\thanks{$^{\mbox{\Letter}}$Yinfeng Yu is the corresponding author(E-mail: yuyinfeng@xju.edu.cn).}%
\\
$^1$Joint Research Laboratory for Embodied Intelligence, Xinjiang University\\
$^2$Joint International Research Laboratory of Silk Road Multilingual Cognitive Computing, Xinjiang University\\
$^3$School of Computer Science and Technology, Xinjiang University, Urumqi 830017, China%
}

\maketitle

\begin{abstract}
In Audio-Visual Navigation (AVN), agents must locate sound sources in unseen 3D environments using visual and auditory cues. However, existing methods often struggle with generalization in unseen scenarios, as they tend to overfit to semantic sound features and specific training environments. 
To address these challenges, we propose the \textbf{Binaural Difference Attention with Action Transition Prediction (BDATP)} framework, which jointly optimizes perception and policy. Specifically, the \textbf{Binaural Difference Attention (BDA)} module explicitly models interaural differences to enhance spatial orientation, reducing reliance on semantic categories. Simultaneously, the \textbf{Action Transition Prediction (ATP)} task introduces an auxiliary action prediction objective as a regularization term, mitigating environment-specific overfitting. 
Extensive experiments on the Replica and Matterport3D datasets demonstrate that BDATP can be seamlessly integrated into various mainstream baselines, yielding consistent and significant performance gains. Notably, our framework achieves state-of-the-art Success Rates across most settings, with a remarkable absolute improvement of up to 21.6 percentage points in Replica dataset for unheard sounds. These results underscore BDATP's superior generalization capability and its robustness across diverse navigation architectures.
\end{abstract}

\begin{IEEEkeywords}
Embodied AI, Audio-Visual Navigation, Reinforcement Learning, Binaural Difference Attention, Action Transition Prediction
\end{IEEEkeywords}

\section{Introduction}
\label{sec:intro_rw}

With the rapid development of artificial intelligence~\cite{yu2025dope,yu2023echo,jaquier2025transfer,yu2023measuring} and robotics~\cite{liu2024caven,durante2024agent,wu2024vision}, autonomous navigation for embodied agents has received growing attention in both virtual and real-world environments~\cite{wen2025zero, iftikhar2024artificial,yu2021weavenet}. Audio-Visual Navigation (AVN) is a representative multimodal task in which an agent is required to locate and reach a sound-emitting target using only egocentric visual observations and binaural audio signals in previously unseen environments~\cite{chen2020soundspaces, gan2020look,yusound,yu2025dynamic}. 
Despite recent progress~\cite{shi2025towards,chen2021waypoints}, existing AVN methods still suffer from limited generalization ability~\cite{wang2023learning,Yu_2022_BMVC,li2025audio}. Their performance often degrades significantly when evaluated on \textit{unheard} sound categories or \textit{unseen} spatial layouts. This limitation reveals fundamental challenges in learning robust multimodal representations and navigation policies that can generalize beyond training conditions~\cite{wu2024embodied,zhao2025audio}.

\begin{figure}[t]
\centering
\includegraphics[scale=0.443]{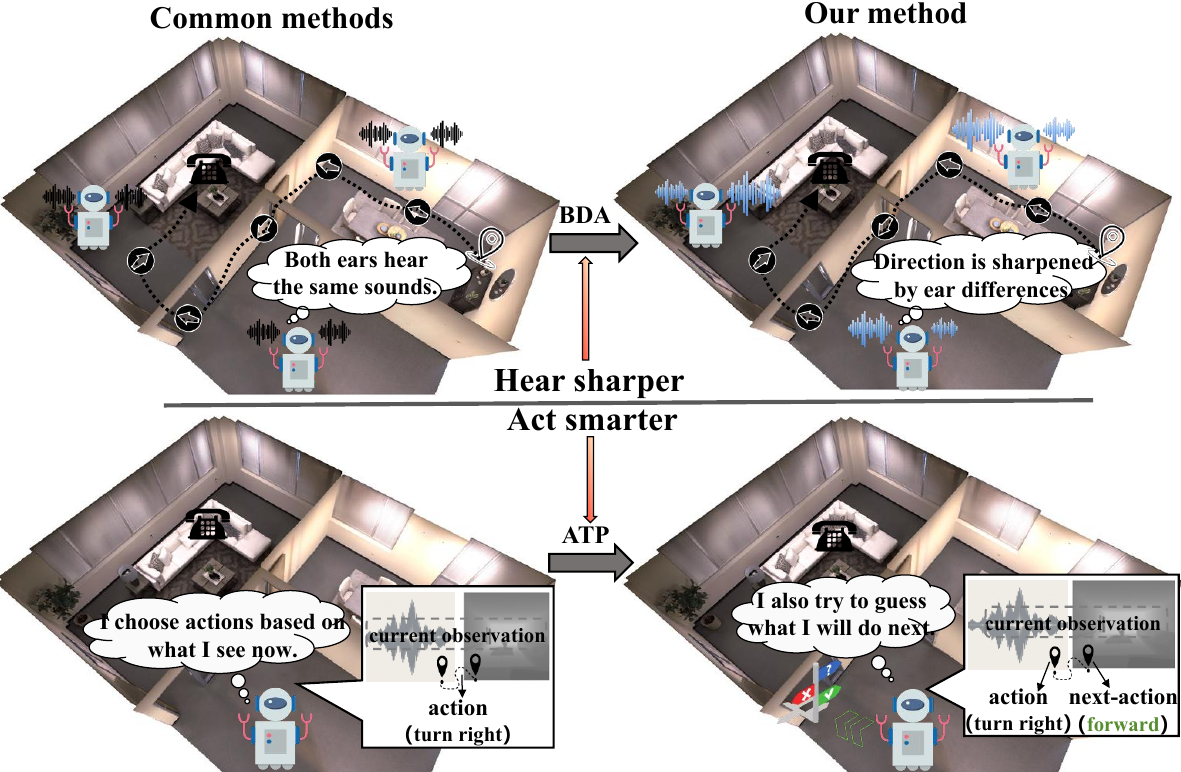}
\caption{Comparison between our BDATP and conventional AVN methods.}
\label{figure1}
\end{figure}

We identify two key factors underlying this issue. First, many audio representations implicitly entangle semantic content with spatial information~\cite{yu2025dgfnet}. While sound semantics may aid recognition, they are often unreliable under category shifts and can obscure universal localization cues~\cite{wang2023learning}, making directional inference unstable for unheard sounds. Second, reinforcement learning-based navigation policies are prone to overfitting to the dynamics and geometries of training environments~\cite{roman2025generating}, resulting in inefficient or unstable behaviors—such as oscillation or backtracking—when deployed in novel scenes.

To overcome these challenges, we propose a unified framework that enhances both perception and policy learning. From a perceptual perspective, the core idea is to prioritize binaural spatial differences over sound semantics, enabling the agent to \textit{Hear Sharper} by focusing on directionally informative cues that generalize across sound categories. From a policy perspective, encouraging temporal consistency in action transitions regularizes navigation behavior, allowing the agent to \textit{Act Smarter} with smoother and more stable trajectories in unseen environments. As illustrated in Fig.~\ref{figure1}, this joint design fundamentally differs from conventional AVN pipelines by explicitly improving spatial perception and policy robustness in a complementary manner.

Extensive experiments on the Replica~\cite{replica19arxiv} and Matterport3D~\cite{Matterport3D} datasets demonstrate that the proposed framework achieves significant improvements over existing methods across comprehensive benchmarks, particularly in challenging zero-shot generalization settings. Our main contributions are summarized as follows:
\begin{itemize}
    \item We propose a binaural spatial perception mechanism that emphasizes interaural differences, enabling agents to \textit{Hear Sharper} under unheard sound categories.
    \item We introduce a policy regularization strategy that promotes temporally consistent action transitions, allowing agents to \textit{Act Smarter} in unseen environments.
    \item We conduct comprehensive evaluations demonstrating strong generalization performance across novel sounds and unseen spatial layouts.
\end{itemize}

\begin{figure*}[b]
\centering
\includegraphics[scale=0.7]{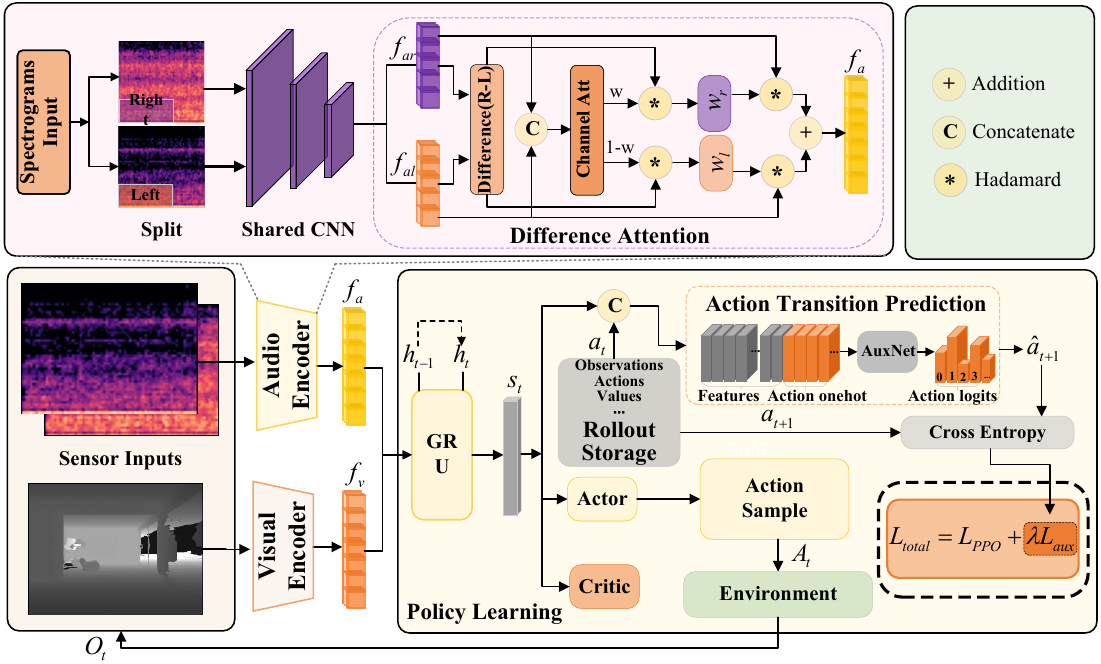}
\caption{BDATP framework overview: separate encoding of visual and auditory inputs, with pink (BDA) and yellow (ATP) modules.}
\label{figure2}
\end{figure*}

\section{Related Work}
\label{sec:related_work}

Audio-Visual Navigation (AVN) has been extensively studied as a multimodal embodied task that integrates auditory and visual cues for target-driven navigation. Early work, such as SoundSpaces~\cite{chen2020soundspaces}, established realistic acoustic simulation environments and demonstrated the feasibility of end-to-end reinforcement learning for sound-guided navigation. Subsequent approaches introduced architectural and algorithmic enhancements to improve navigation efficiency and robustness, including hierarchical planning with intermediate waypoints~\cite{chen2021waypoints}, semantic memory for intermittent sound sources~\cite{chen2021semantic}, and omnidirectional perception for richer environmental awareness~\cite{chen2023omnidirectional}. 

More recent generalization remains a central challenge in AVN, particularly when agents are evaluated on unheard sound categories or unseen environments. Several works attempt to reduce reliance on sound semantics by introducing auxiliary objectives or learning more abstract audio representations~\cite{wang2023learning}. However, many existing methods still depend on implicit feature learning through black-box encoders, which may struggle to disentangle spatial cues from semantic content under distribution shifts. From the policy perspective, reinforcement learning-based navigation agents are known to overfit to the specific dynamics and layouts of training environments, leading to unstable trajectories in novel scenes~\cite{chen2021waypoints, chen2023omnidirectional,zhang2025advancing}. While prior efforts have explored transfer learning or pre-training strategies to improve robustness, learning environment-agnostic and temporally consistent navigation behaviors remains an open problem in audio-visual navigation.

\begin{table*}[t]
\centering
\caption{Performance comparison on Replica and Matterport3D datasets}
\label{tab:main_results}
\begin{tabular}{l ccc ccc ccc ccc}
\toprule
\multirow{2}{*}{\textbf{Method}} & \multicolumn{6}{c}{\textbf{Replica}} & \multicolumn{6}{c}{\textbf{Matterport3D}} \\
\cmidrule(lr){2-7} \cmidrule(lr){8-13}
& \multicolumn{3}{c}{Heard} & \multicolumn{3}{c}{Unheard} & \multicolumn{3}{c}{Heard} & \multicolumn{3}{c}{Unheard} \\
& SR$\uparrow$ & SPL$\uparrow$ & SNA$\uparrow$ & SR$\uparrow$ & SPL$\uparrow$ & SNA$\uparrow$ & SR$\uparrow$ & SPL$\uparrow$ & SNA$\uparrow$ & SR$\uparrow$ & SPL$\uparrow$ & SNA$\uparrow$ \\
\midrule
Random Agent        & 18.5 & 4.9  & 1.8  & 18.5 & 4.9  & 1.8  & 9.1  & 2.1  & 0.8  & 9.1  & 2.1  & 0.8  \\
Direction Follower  & 72.0 & 54.7 & 41.1 & 17.2 & 11.1 & 8.4  & 41.2 & 32.3 & 23.8 & 18.0 & 13.9 & 10.7 \\
SAVi~\cite{chen2021semantic}                    & 54.0 & 45.1 & 30.8 & 33.9 & 27.5 & 17.2 & 40.3 & 29.1 & 13.0 & 29.5 & 20.4 & 9.6  \\
Dav-NaV~\cite{younes2023catch}                 & 85.1 & 72.6 & 54.0 & 58.5 & 45.6 & 33.4 & 82.9 & 61.9 & 46.8 & 55.3 & 42.4 & 31.6 \\
SA2GVAN~\cite{wang2023learning}                & 90.4 & 70.9 & 55.2 & 62.8 & 43.4 & 33.0 & 82.9 & 61.4 & 46.8 & 60.7 & 42.3 & 31.4 \\
ORAN~\cite{chen2023omnidirectional}            & -    & -    & -    & 60.9 & 46.7 & \textbf{36.5} & -    & -    & -    & 59.4 & \textbf{50.8} & \textbf{35.2} \\      
\midrule
AV-NaV~\cite{chen2020soundspaces}              & 88.9 & 64.5 & 44.1 & 47.3 & 34.7 & 14.1 & 66.2 & 44.8 & 27.3 & 33.5 & 21.9 & 10.4 \\
\textbf{AV-NaV + BDATP}                        & 93.1 & 74.5 & 43.9 & 68.6 & 45.0 & 19.4 & 68.7 & 51.7 & 28.2 & 55.1 & 37.9  & 20.1 \\
\midrule
AV-WaN~\cite{chen2021waypoints}                & 90.9 & 70.4 & 52.5 & 52.8 & 34.7 & 27.1 & 82.4 & 55.4 & 42.5 & 56.7 & 40.9 & 30.4 \\
\textbf{AV-WaN + BDATP}                        & \textbf{96.5} & \textbf{79.2} & \textbf{63.5} & \textbf{70.7} & \textbf{49.9} & 34.6 & \textbf{85.4} & \textbf{66.4} & \textbf{52.1} & \textbf{65.4} & 44.0 & 32.7 \\
\bottomrule
\end{tabular}
\end{table*}

\section{Method}
\label{sec:method}
To enhance generalization in AVN, we propose the BDATP framework (Fig. \ref{figure2}), comprising two core components: a \textbf{Binaural Difference Attention (BDA)} module to capture universal interaural spatial cues for robust perception, and an \textbf{Action Transition Prediction (ATP)} auxiliary task to reinforce cross-environment statistical regularities during policy learning.The following subsections detail these components.

\subsection{Feature Extraction and Binaural Difference Attention}
In the feature extraction stage, the visual input is a depth image, and the auditory input is a binaural spectrogram. Similar to AV-NaV, we employ two independent CNN encoders, each consisting of convolutional layers with kernel sizes of $8 \times 8$, $4 \times 4$, and $3 \times 3$, followed by a linear layer with ReLU activations in between. These encoders separately produce the visual feature $f_v \in \mathbb{R}^{512}$ and the audio feature $f_a \in \mathbb{R}^{512}$.

Conventional approaches often concatenate the binaural spectrograms directly, which tends to ignore interaural differences and weakens the agent’s ability to perceive sound direction, especially leading to degraded performance on unseen sound categories~\cite{zhang2025iterative}. To address this, we introduce the BDA module within the audio encoder, which is applied before the linear layer. Specifically, the binaural spectrogram is split into the left and right channels before encoding. A CNN with shared weights encodes them separately, yielding intermediate feature maps $f_{al}$ and $f_{ar}$. We then compute the element-wise difference and concatenate the original channel features:  
\begin{align}
diff &= |f_{ar} - f_{al}| \in \mathbb{R}^{B\times C\times h\times w},\\
f'_a &= \operatorname{Concat}(f_{al}, f_{ar}) \in \mathbb{R}^{B\times 2C\times h\times w}.
\end{align}
The concatenated feature $f'_a$ is projected back from $2C$ channels to $C$ via a $1\times1$ convolution, followed by a Sigmoid activation to obtain channel-wise attention weights:$w  \in (0,1)^{B\times C\times h\times w}$.
Unlike conventional interpolation-based weighting, we explicitly use $diff$ as a modulation term, with $w$ distributing emphasis between the left and right channels:
\begin{align}
w_r &= w \odot diff, \\
w_l &= (1-w)\odot diff, \\
f_a &= f_{al} \odot w_l \;+\; f_{ar} \odot w_r.
\end{align}
Here, $w_r$ and $w_l$ are the difference-weighted coefficients for the right and left channels, and $f_a$ is the fused audio feature.Compared to simple concatenation or weighting, BDA enforces attention to binaural spatial differences while preserving original channel amplitudes. This allows the model to leverage cross-category spatial cues for better generalization to unseen sounds.

\subsection{Policy Learning with Action Transition Prediction}
In reinforcement learning-based AVN, sparse reward signals often cause policies to overfit to training environments, resulting in unstable trajectories and poor generalization to unseen scenes. To mitigate this, we propose the Action Transition Prediction (ATP) auxiliary task. ATP leverages intra-rollout temporal supervision to impose a global statistical regularization across parallel environments. Specifically, it penalizes the discrepancy between the predicted action for the next timestep and the actual action executed by the agent within the same trajectory. By incorporating this prediction error as a regularization term, ATP encourages the policy to prioritize environment-invariant features—extracting cross-modal representations essential for navigation.

Formally, at each timestep $t$, given the current state feature $s_t$ and the action $a_t$ taken at the previous timestep, an auxiliary network predicts the next action $\hat{a}_{t+1}$:  
\begin{equation}
\hat{a}_{t+1} = \mathrm{AuxNet}([s_t; \mathrm{OneHot}(a_t)]),
\end{equation}
where $\text{AuxNet}$ is a two-layer fully-connected network. The cross-entropy loss is computed using the ground-truth action $a_{t+1}$ that was actually performed at the next timestep within the same rollout:
\begin{equation}
\mathcal{L}_{aux} = \frac{1}{(T-1) \cdot B} \sum_{i=1}^{(T-1)\cdot B} - \log \frac{\exp([\hat{a}_{t+1}]_{i, (a_{t+1})_i})}{\sum_{j=1}^{N} \exp([\hat{a}_{t+1}]_{i,j})},
\end{equation}
where $[\hat{a}_{t+1}]_{i,j}$ represents the predicted logits for sample $i$ and action $j$, $(a_{t+1})_i \in \{0, \dots, N-1\}$ is the ground-truth action index, $B$ is the number of parallel environments (batch size), and $T$ is the rollout length. Specifically,the value of $N$ denotes the size of the action space, which varies depending on the navigation backbone:For AV-NaV, $N=4$ corresponding to discrete low-level actions (Forward, Left, Right, Stop).For AV-WaN, $N=81$ as the model predicts intermediate waypoints across a $9 \times 9$ spatial action map.

Finally, within the PPO framework, the auxiliary loss is added as a regularization term to the policy loss:
\begin{equation}
\mathcal{L}_{total} = \mathcal{L}_{PPO} + \lambda \mathcal{L}_{aux},
\end{equation}
where $\mathcal{L}_{PPO}$ includes the clipped surrogate loss, value function loss, and entropy regularization. We set $\lambda=0.1$ to balance auxiliary supervision with reward-driven optimization.

The key insight of ATP is that while specific states $s_t$ vary across environments, the action transitions reflect consistent navigational regularities. For instance, agents generally exhibit similar turning probabilities when encountering obstacles or lateral sound gradients regardless of the specific room geometry. Although parallel rollouts yield diverse state-action pairs, they all encode these common underlying transition dynamics. By jointly optimizing the transition prediction loss across a batch of rollouts, ATP enforces a shared statistical constraint, strengthening the policy’s sensitivity to generalizable cues and improving transferability to novel environments.

\section{Experiments and Analysis}
To assess the effectiveness and generalization of our method in audiovisual navigation, we conduct experiments on two realistic 3D simulation datasets and compare with state-of-the-art baselines. These baselines include heuristic methods (\textbf{Random} and \textbf{Direction Follower}), the fundamental RL-based \textbf{AV-NaV}~\cite{chen2020soundspaces}, the hierarchical waypoint-based \textbf{AV-WaN}~\cite{chen2021waypoints}, the semantic-aware \textbf{SAVi}~\cite{chen2021semantic}, the robust spatial-fusing \textbf{Dav-NaV}~\cite{younes2023catch}, the omnidirectional navigator \textbf{ORAN}~\cite{chen2023omnidirectional}, and the semantic-agnostic \textbf{SA2GVAN}~\cite{wang2023learning}. This section details the experimental setup, quantitative and qualitative results, and trajectory analysis.

\subsection{Experimental Setup}
Experiments are conducted in the SoundSpaces~\cite{chen2020soundspaces} simulation platform using two real-world 3D scene datasets: Replica~\cite{replica19arxiv} and Matterport3D~\cite{replica19arxiv}. The Replica environments are relatively small (average area 47.24 m\(^2\)), while Matterport3D contains larger and more complex environments (average area 517.34 m\(^2\)). To demonstrate the effectiveness of our framework, we validate our method on two representative baselines, AV-NaV~\cite{chen2020soundspaces} and AV-WaN~\cite{chen2021waypoints}.

Navigation performance is evaluated by the following metrics~\cite{anderson2018evaluation}:  
(1) \textbf{SR} (Success Rate): measures whether the agent successfully reaches the target location.  
(2) \textbf{SPL} (Success weighted by Path Length): combines success rate with path efficiency.  
(3) \textbf{SNA} (Success weighted by Number of Actions): assesses the economy of agent behavior by weighing success against the ratio of optimal to actual execution steps.

We consider two evaluation settings based on sound category exposure: \textbf{Heard} and \textbf{Unheard}, both involving multiple sound types. In the Heard setting, test sound categories appear in training but test scenes are unseen. In the Unheard setting, both sound categories and scenes are unseen.

\begin{figure*}[tb]
\centering
\includegraphics[scale=0.55]{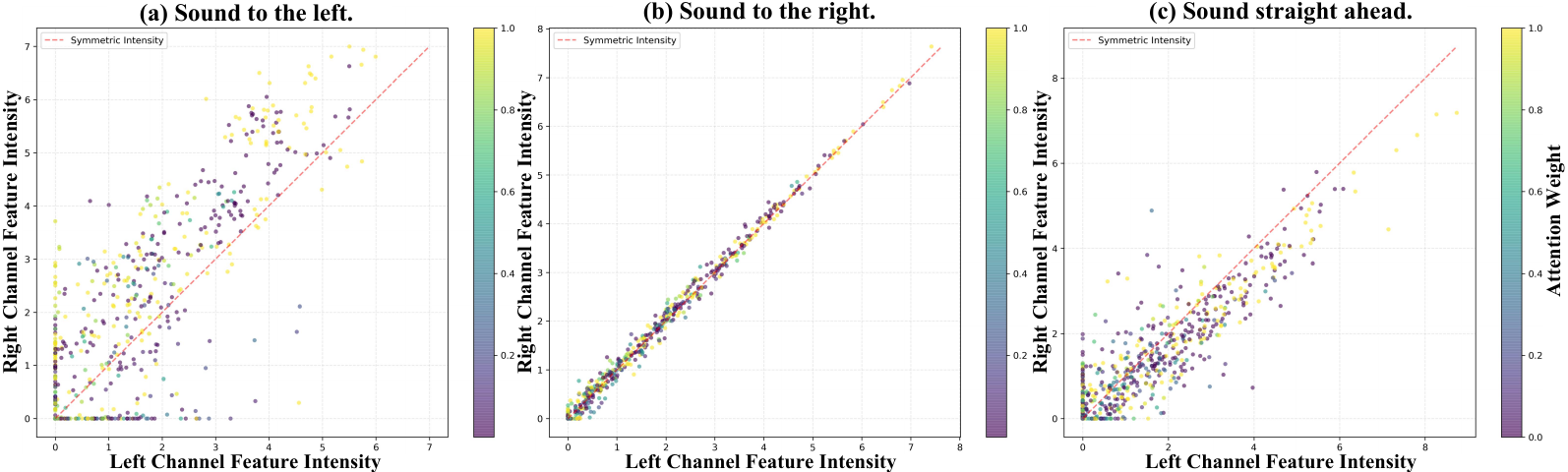}
\caption{Feature relationship visualization of the BDA module across three spatial scenarios.}
\label{figure3}
\end{figure*}

\subsection{Results and Analysis} 

\subsubsection{Quantitative Results} 

Table~\ref{tab:main_results} reports the performance comparison on Replica and Matterport3D datasets. Overall, our BDATP framework consistently enhances the navigation capabilities of mainstream baselines, achieving state-of-the-art results across most metrics. 

Specifically, on the Replica dataset, integrating BDATP with AV-WaN yields the highest performance. In the \textit{Heard} setting, BDATP improves the SR of AV-WaN from 90.9\% to 96.5\% and SPL from 70.4\% to 79.2\%. The advantage is even more pronounced in the challenging \textit{Unheard} setting, where AV-WaN + BDATP achieves an SR of 70.7\% and an SPL of 49.9\%, outperforming the original AV-WaN baseline by 17.9\% and 15.2\%, respectively. Similarly, when applied to the AV-NaV baseline, BDATP boosts the \textit{Unheard} SR from 47.3\% to 68.6\%, a substantial improvement of 21.3 percentage points. These results validate that the BDA module effectively captures essential binaural spatial cues, while the ATP task successfully regularizes the policy for better generalization to novel sound categories.

On the larger and more acoustically complex Matterport3D dataset, BDATP maintains robust performance gains. The AV-WaN + BDATP configuration achieves the best SR of 85.4\% in the \textit{Heard} setting and 65.4\% in the \textit{Unheard} setting. Notably, even when integrated with the simpler AV-NaV architecture, BDATP boosts the \textit{Unheard} SR by 21.6 percentage points (from 33.5\% to 55.1\%). This consistent improvement across different architectures and dataset scales underscores the plug-and-play nature of our framework. Furthermore, the significant increase in SNA across most experimental settings confirms that our method not only reaches the goal more frequently but also does so with fewer redundant actions, resulting in more efficient and stable navigation trajectories.

\subsubsection{Qualitative Analysis}

\begin{figure*}[htbp]
\centering
\includegraphics[scale=0.4]{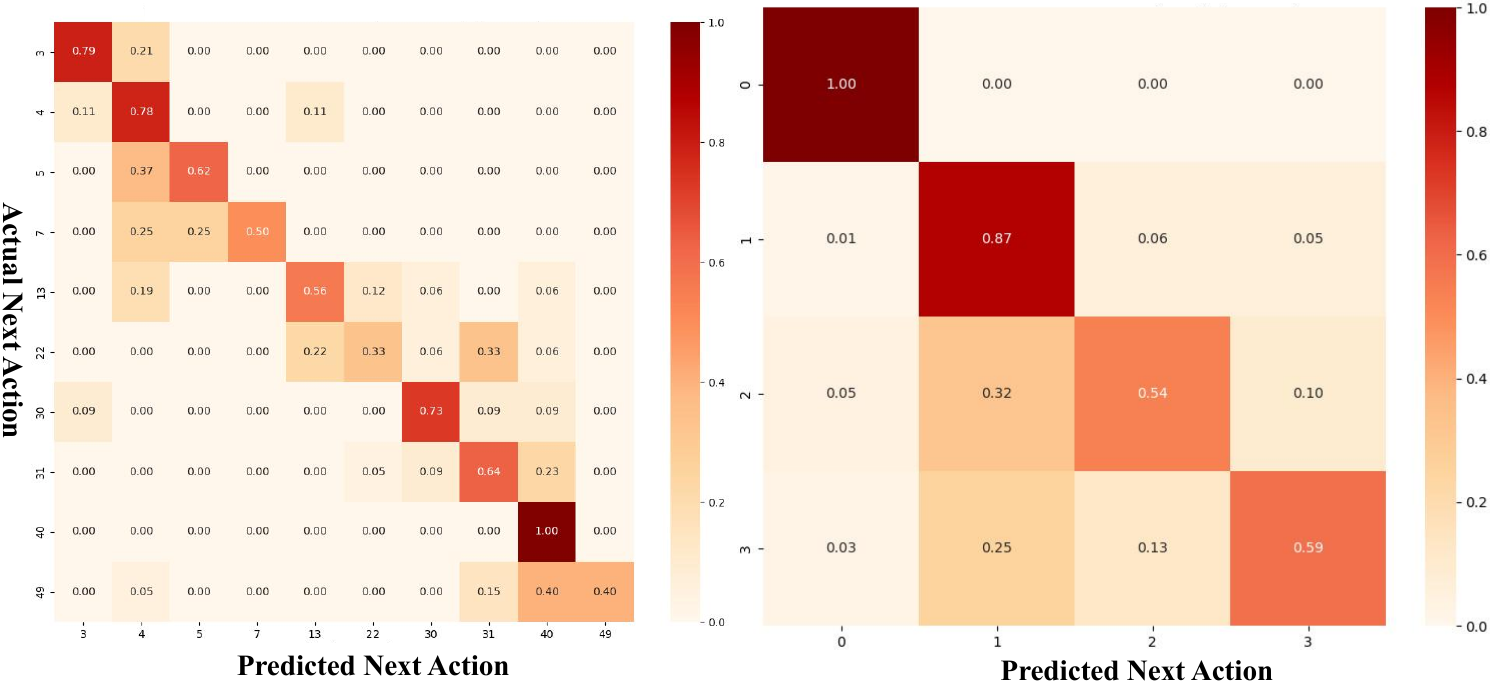}
\caption{Action Transition Matrices of ATP.Left: Top-10 most probable transitions on AV-WaN; Right: Full transitions on AV-Nav.}
\label{figure4}
\end{figure*}

\begin{figure*}[b]
\centering
\includegraphics[scale=0.65]{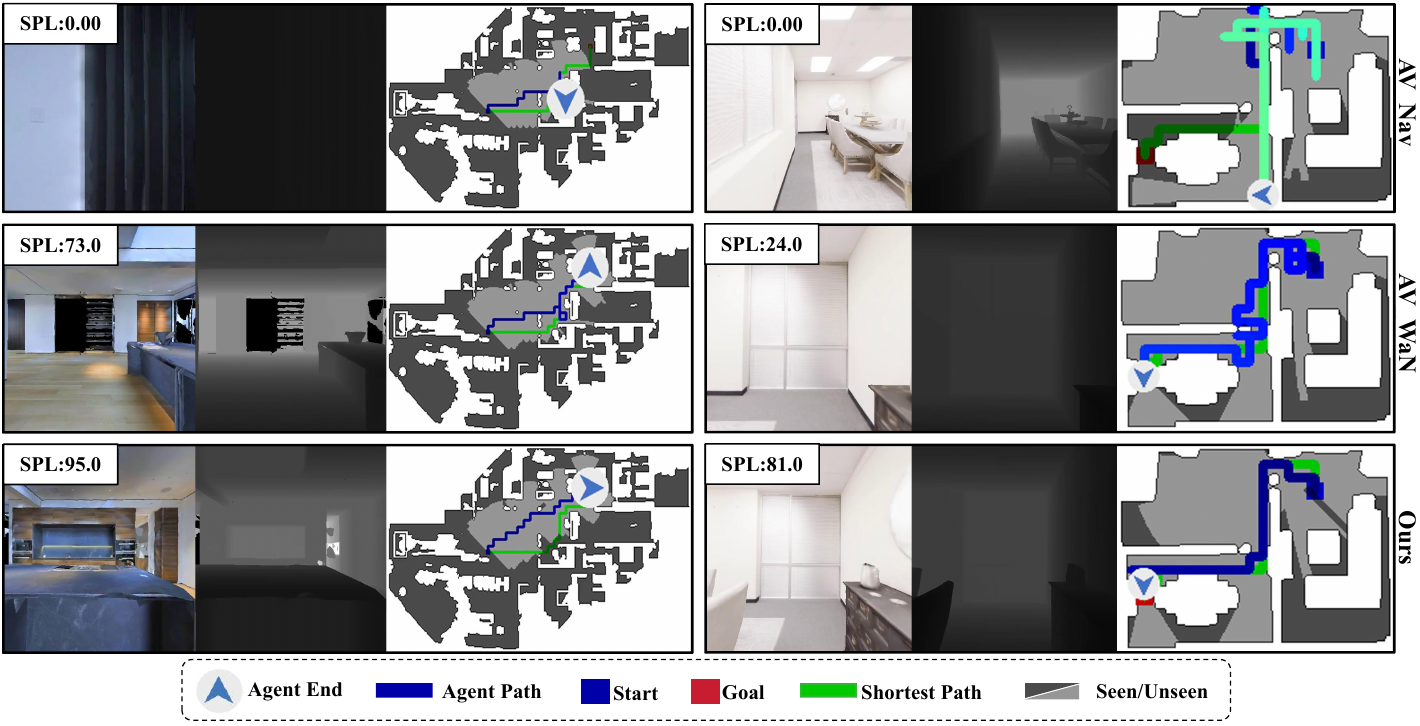}
\caption{Trajectory comparison in unseen/unheard settings: Matterport3D (left) and Replica (right).}
\label{figure5}
\end{figure*}

To better understand BDATP’s generalization ability, we analyze binaural perception, action transitions, and navigation trajectories. As shown in Fig.~\ref{figure3}, binaural feature distributions under directional acoustic conditions deviate from the diagonal ($y=x$), indicating strong sensitivity to interaural differences and enabling robust spatial localization even for unheard sound categories. Fig.~\ref{figure4} shows that the ATP auxiliary task produces a clear diagonal pattern in action transition matrices, reflecting high consistency with ground truth and reduced transition ambiguity. As illustrated in Fig.~\ref{figure5}, BDATP generates smoother and more stable trajectories on both Matterport3D and Replica datasets in the Unheard setting, while baseline methods often fail or exhibit inefficient behaviors. These results demonstrate that enhancing spatial perception and action transition reliability leads to more efficient and robust navigation toward novel sound sources.

\subsubsection{Ablation Study} 

\begin{table}[t]
\centering
\caption{Ablation study results on Replica datasets}
\label{tab:ablation}
\begin{tabular}{l ccc c} 
\toprule
\textbf{Model} & \multicolumn{2}{c}{\textbf{Heard}} & \multicolumn{2}{c}{\textbf{Unheard}} \\
\cmidrule(lr){2-5} 
 & SR$\uparrow$ & SPL$\uparrow$ & SR$\uparrow$ & SPL$\uparrow$ \\
\midrule
w/o BDA and ATP & 88.9 & 64.5 & 47.3 & 34.7 \\
w/o ATP         & 90.2 & 74.0 & 66.2 & 44.4 \\
w/o BDA        & 92.2 & 72.9 & 63.4 & 44.3 \\
\midrule
\textbf{AV-NaV + BDATP}    & \textbf{93.1} & \textbf{74.5} & \textbf{68.6} & \textbf{45.0} \\
\bottomrule
\end{tabular}
\end{table}

\begin{table}[htbp]
\centering
\small
\caption{Ablation study of the auxiliary task loss weight $\lambda$ on Replica}
\label{tab:lambda_ablation}
\begin{tabular}{l ccc c}
\toprule
\multirow{2}{*}{\textbf{Weight} $\lambda$} & \multicolumn{2}{c}{\textbf{Heard}} & \multicolumn{2}{c}{\textbf{Unheard}} \\
\cmidrule(lr){2-3} \cmidrule(lr){4-5}
 & SR$\uparrow$ & SPL$\uparrow$ & SR$\uparrow$ & SPL$\uparrow$ \\
\midrule
$\lambda = 0$ & 88.9 & 64.5 & 47.3 & 34.7 \\
$\lambda = 0.001$ & 90.2 & 66.3 & 57.1 & 39.2 \\
$\lambda = 0.01$ & 88.9 & 68.8 & 62.9 & 39.1 \\
\midrule
$\lambda = 0.1$(Ours)   & \textbf{92.2} & \textbf{72.9} & \textbf{63.4} & \textbf{44.3} \\
\bottomrule
\end{tabular}
\end{table}

We evaluate the contributions of BDA and ATP on the AV-NaV architecture using the Replica dataset (Table~\ref{tab:ablation}). Removing both modules causes a clear performance drop, especially in the \textit{Unheard} setting, where SR falls to 47.3\%, highlighting the need for both spatial perception and policy learning.
Without BDA, the \textit{Unheard} SR decreases from 68.6\% to 63.4\%, showing that modeling interaural differences is crucial for generalizing to novel sound categories. Removing ATP mainly affects stability and efficiency: while \textit{Heard} performance remains strong, \textit{Unheard} SR drops to 66.2\% and SPL is consistently lower. Overall, BDA provides reliable directional cues, while ATP improves policy robustness through more consistent action transitions.

We further analyze the sensitivity to the auxiliary task loss weight $\lambda$ based on the AV-NaV architecture without BDA (Table~\ref{tab:lambda_ablation}), where $\lambda=0$ represents the vanilla AV-NaV baseline.Increasing $\lambda$ to 0.1 consistently enhances performance, particularly boosting the Unheard SR from 47.3\% to 63.4\%. This 16.1 percentage point improvement confirms that the ATP task provides crucial policy regularization.

\section{Conclusion}
This paper presents \textbf{BDATP}, a framework designed to enhance generalization in Audio-Visual Navigation. By integrating \textbf{BDA} for robust spatial perception and \textbf{ATP} for shared statistical regularization of action transitions, our method effectively mitigates overfitting to specific sound categories and environments. Extensive evaluations on Replica and Matterport3D confirm that BDATP serves as a versatile plug-and-play method, consistently improving navigation performance across multiple baselines. Future work will extend this framework to dynamic sound sources and multi-agent coordination in real-world acoustic environments.

\section*{ACKNOWLEDGMENT}

This research was financially supported by the National Natural Science Foundation of China (Grant No.: 62463029).

\bibliographystyle{IEEEtran}
\bibliography{refs}  

\end{document}